**Synchrotron X-ray Diffraction Study of BaFe$_2$As$_2$ and CaFe$_2$As$_2$ at High Pressures up to 56 GPa: Ambient and Low-Temperatures Down to 33 K**

R. Mittal, S. K. Mishra and S. L. Chaplot
*Solid State Physics Division, Bhabha Atomic Research Centre, Trombay, Mumbai 400 085, India*

S. V. Ovsyannikov[1], E. Greenberg[1,2], D. M. Trots[1] and L. Dubrovinsky[1]
[1]*Bayerisches Geoinstitut, University Bayreuth, Universitatsstrasse 30, D-95440 Bayreuth, Germany*
[2]*School of Physics & Astronomy, Tel-Aviv University, Israel*

Y. Su[3] and Th. Brueckel[3,4]
[3]*Juelich Centre for Neutron Science, IFF, Forschungszentrum Juelich, Outstation at FRM II, Lichtenbergstr. 1, D-85747 Garching, Germany*
[4]*Institut fuer Festkoerperforschung, Forschungszentrum Juelich, D-52425 Juelich, Germany*

S. Matsuishi[5] and H. Hosono[5]
[5]*Frontier Research Center, Tokyo Institute of Technology, 4259 Nagatsuta-cho, Midori-ku, Yokohama 226-8503, Japan*

G. Garbarino[6]
[6]*European Synchrotron Radiation Facility, BP 220, 38043 Grenoble, France*

**Abstract**

We report high pressure powder synchrotron x-ray diffraction studies on $M$Fe$_2$As$_2$ ($M$=Ba, Ca) over a range of temperatures and pressures up to about 56 GPa using a membrane diamond anvil cell. A phase transition to a collapsed tetragonal phase is observed in both compounds upon compression. However, at 300 (33) K in the Ba-compound the transition occurs at 26 (29) GPa, which is a much higher pressure than 1.7 (0.3) GPa at 300 (40) K in the Ca-compound, due to its larger volume. It is important to note that the transition in both compounds occurs when they are compressed to almost the same value of the unit cell volume and attain similar $c_t/a_t$ ratios. We also show that the FeAs$_4$ tetrahedra are much less compressible and more distorted in the collapsed tetragonal phase than their nearly regular shape in the ambient pressure phase. We present a detailed analysis of the pressure dependence of the structures as well as equation of states in these important BaFe$_2$As$_2$ and CaFe$_2$As$_2$ compounds.

## I. Introduction

The iron-based superconductors $Ln$FeAs(O$_{1-x}$F$_x$) ($Ln$=Lanthanides) and oxygen-free doped $M$Fe$_2$As$_2$ ($M$= Ba, Ca, Sr, Eu), LiFeAs, FeSe, and SrFeAsF have stimulated a great deal of activity [1-33] on superconductors derived from antiferromagnetic parent compounds. This novel class of materials, besides existing cuprate-based high-$T_c$ superconductors, provides yet another system for exploring the interplay between superconductivity and antiferromagnetism. The parent $M$Fe$_2$As$_2$ compounds containing FeAs layers exhibit structural [5] and magnetic phase transitions [5-7] associated with Fe moments. For example, BaFe$_2$As$_2$ undergoes structural (tetragonal to orthorhombic) and magnetic (paramagnetic to antiferromagnetic (AF)) phase transitions simultaneously at $T_s$ ~140 K. The chemical substitutions of Ba for K [2] and Fe for Co [28], and pressure application suppress the AF transition [2-4], resulting in the appearance of superconductivity. The spin fluctuations [25, 29] are suggested to play an important role in establishing the superconducting ground state. Extensive studies of phonon dynamics [8-11] suggest that it is unlikely that the superconductivity in iron pnictides is due to simple electron-phonon coupling. Thus, superconductivity in iron arsenide materials is associated with the lattice distortion and suppression of magnetic ordering, indicating a strong competition between the structural distortions, magnetic ordering and superconductivity. The detailed interplay between the crystal structure, magnetic ordering, and superconductivity is hardly understood up to now, which is to some extent due to the lack of precise structural data at high pressures. In addition to this, an important clue to the mechanism of superconductivity should be provided by high-pressure experiments on a stoichiometric sample since the application of pressure introduces no disorder. However, fundamental problems remain to be solved, as the appearance of pressure-induced superconductivity in $M$Fe$_2$As$_2$ is highly sensitive [4, 21, 22] to pressure homogeneity.

In CaFe$_2$As$_2$, detailed neutron- and x-ray diffraction analysis shows [12, 13] that the system undergoes a first-order phase transition from a magnetic orthorhombic to a nonmagnetic "collapsed" tetragonal phase at $P$=0.3 GPa, $T$=40 K. The inclusion of only a small amount of tetragonal phase gives rise to a spurious superconductivity in the magnetic orthorhombic phase. A superconducting phase has been found [30, 31, 33] in a collapsed tetragonal structure of CaFe$_2$As$_2$ with the disappearance of magnetism. However, the possible appearance of superconductivity in the collapsed tetragonal phase of CaFe$_2$As$_2$ is presently under debate [4]. These measurements were performed in a limited pressure range (< 1.8 GPa) [12, 13, 30, 31, 33]. In the case of BaFe$_2$As$_2$, the superconducting dome evolves [2, 3, 14, 21, 32] in a gradual way with the change of pressure (chemical/external) and superconductivity has been reported to exist in the orthorhombic structure, suggesting a coexistence of superconductivity and magnetic order. Although great efforts have been devoted to understand the relationship between magnetism, lattice instability and superconductivity, there are still some discrepancies and unclear issues. Recently, high-pressure measurements carried out [16] for $EuFe_2As_2$ at 300 K show a tetragonal to collapsed tetragonal phase transition at about 8 GPa.

The increase of superconducting transition temperature $T_c$ by application of external pressure suggests [16] that the lattice contraction is effective in the enhancement of $T_c$ of iron-based superconductors. The chemical and external pressures are found to play [14] similar roles in leading to superconductivity, while also suppressing the tetragonal-to-orthorhombic phase transition and reducing the As–Fe–As bond angle and Fe–Fe distance. In particular, $T_C$ is found to increase as the As-Fe-As bond angles tend towards the ideal tetrahedral value of $109.47^0$.

First-principles band structure calculations have been used [25] to understand the relationship between the crystal structure, charge doping and magnetism in $BaFe_2As_2$ and $CaFe_2As_2$. Recently, ab initio molecular dynamics [26, 27] calculations have also been used to investigate the electronic and structural phase transformation of $BaFe_2As_2$ and $CaFe_2As_2$ under pressure. The authors report that the structural phase transition from orthorhombic to tetragonal symmetry is accompanied by a magnetic phase transition in all the compounds while the nature of the transition is different for the two systems.

It is interesting to note that most of the high-temperature superconductors have a layered structure. It is also important to know how pressure changes the interaction between the layers, which can be directly determined from the measurement of the compressibility of the lattice parameters. We have investigated the pressure effects on $CaFe_2As_2$ and $BaFe_2As_2$ using synchrotron x-ray diffraction technique. For $CaFe_2As_2$, Rietveld analysis of powder synchrotron data clearly suggests that it undergoes a phase transformation from a tetragonal to a collapsed tetragonal phase via an orthorhombic phase at very low pressures as is well known in the literature. Powder synchrotron diffraction data does not show any appreciable change as pressure is increased up to ~ 37 GPa at 40 K. The compound remains in the collapsed tetragonal phase. On the other hand, $BaFe_2As_2$ undergoes a tetragonal to orthorhombic phase transition around 130 K at about 1.3 GPa and remains in that phase upon lowering the temperature down to 33 K. An increase in pressure (at 33 K) results in a broadening of the diffraction profile at about 29 GPa. Detailed Rietveld analysis of the diffraction data clearly indicates a coexistence of orthorhombic and collapsed tetragonal phases above 29 GPa at 33 K. The phase fraction of collapsed tetragonal phase increases with pressure. Upon heating from 33 to 300 K (at ~34 GPa), we found that $BaFe_2As_2$ is fully transformed into the collapsed tetragonal phase. Finally, at 300 K, while lowering the pressure, BaFe$_2$As$_2$ is transformed again back to the tetragonal phase at about 20 GPa. In addition to this, high pressure data was also collected for both compounds up to about 56 GPa at 300 K, indicates a tetragonal to collapsed tetragonal phase transition in BaFe$_2$As$_2$ at about 26 GPa. The details of the measurements are given in section II. The results and discussion, and conclusions are presented in sections III and IV, respectively.

**II. Experimental**

The polycrystalline samples of $M$Fe$_2$As$_2$ ($M$: Ba, Ca) were prepared by heating stoichiometric mixtures of the corresponding purified elements. The x-ray powder diffraction measurements were carried out at the ID-27 beam line at the European Synchrotron Radiation Facility (ESRF, Grenoble, France). An applied pressure was generated by membrane diamond anvil cells (DACs). A powdery sample of ~ 30-40 µm in the diameter and 10 µm in the thickness was situated in the centre of diamond anvil's tip. A pressure was generated into a stainless steel gasket pre-indented to a 40-50 µm thickness, with a central hole of 150 µm in diameter filled with helium as pressure transmitting medium. The pressure was determined using the shift of the fluorescence line of the ruby. The wavelength of the x-ray (0.3738 Å) was selected and determined using the a Si(111) monochromator and the iodine K-edge. Then the sample to image plate (MAR345) detector distance was refined using the diffraction data of Si. A continuous helium flow CF1200 DEG Oxford cryostat was used to cool down the DAC. Special care was taken to obtain stable temperature and pressure conditions prior to each acquisition. The precision and accuracy of the temperature measurement is better than 0.1 K and 0.2 K respectively. In the first cycle, the BaFe$_2$As$_2$ and CaFe$_2$As$_2$ sample was first cooled to 33 K and 40 K, respectively, and then pressure was increased to ~35 GPa. At 35 GPa, the temperature was slowly increased to 300 K. Finally, pressure was released at 300 K. Another set of measurements was carried out at ambient temperature (300K) for both the samples, where data was collected up to about 56 GPa and 51 GPa for BaFe$_2$As$_2$ and CaFe$_2$As$_2$, respectively. Typical exposure times of 20 seconds were employed for the measurements.

The 2-dimensional powder images were integrated using the program FIT2D [34] to yield intensity *vs.* 2θ plot. The diffraction patterns have preferred orientation along [103] for BaFe$_2$As$_2$ at 300 K. On the other hand, CaFe$_2$As$_2$ showed preferred orientation along [213] and [200] at 40 K and 300 K, respectively. The structural refinements were performed using the Rietveld refinement program FULLPROF [35]. In all the refinements, the background was defined by a sixth order polynomial in 2θ. A Thompson-Cox-Hastings pseudo-Voigt with axial divergence asymmetry function was chosen to define the profile shape for the powder synchrotron diffraction peaks. The scale factor, background, half-width parameters along with mixing parameters, lattice parameters and positional coordinates, were refined.

### III. Results and discussion

The powder synchrotron x-ray diffraction measurements for $MFe_2As_2$ (M=Ba, Ca) at ambient conditions confirmed a single-phase sample consistent with published reports [2, 13]. The data was collected for $MFe_2As_2$ (M=Ba, Ca) over a wide range of temperatures and pressures up to 56 GPa. The pressure-temperature conditions for measurement in various compression cycles for $BaFe_2As_2$ and $CaFe_2As_2$ are shown in Fig. 1(a) and (b), respectively.

### A. High pressure phase stability of $BaFe_2As_2$

Typical angle dispersive powder X-ray diffraction data collected for $BaFe_2As_2$ at various pressures at 300 K and 33 K are shown in Figs. 2 and 3. Figure 2 (a) shows a portion of the diffraction patterns of $BaFe_2As_2$ at selected pressures at 300 K during the pressure increase cycle. The diffraction profiles show dramatic changes with pressure. The Bragg peaks around $2\theta=10^o$ and $13^o$, come close to each other with increasing pressure and finally merge at about 26 GPa. On further increase of pressure, the Bragg peaks are again well separated for pressures above 34 GPa. Detailed Rietveld refinement of the powder diffraction data shows that diffraction patterns at 300 K could be indexed using the tetragonal structure (space group *I4/mmm*) up to 56 GPa. The fit between the observed and calculated profiles is quite satisfactory and some of them are shown in Figure 2(b).

Earlier energy dispersive X-ray diffraction carried out up to 22 GPa indicated [15] a tetragonal to orthorhombic phase transition for $BaFe_2As_2$ at about 17 GPa at 300 K using a structure free model. In view of this, we have also carried out Rietveld refinement of powder synchrotron data at 22 GPa (similar to ref. 15] with an orthorhombic space group *Fmmm*. We have not found any substantial improvement in $\chi^2$ with orthorhombic phase refinement. The differences in orthorhombic $a_o$ and $b_o$ lattice parameter is found to be less than 0.5%. However Ref. [15] reported a difference of more than 3% in the values of $a_o$ and $b_o$ lattice parameters. Thus, we refined the powder diffraction data using a tetragonal phase and believe that it undergoes a similar phase transition [13] as observed for $CaFe_2As_2$.

In another set of measurements, diffraction patterns were collected while lowering the temperature of the DAC at about 1 GPa. Detailed Rietveld analysis of the diffraction data reveals that it undergoes a tetragonal to orthorhombic phase transition near 130 K at about 1.3 GPa as shown in Fig. 3. The structure was found to remain in the orthorhombic phase (Fig. 4(a)) down to 33 K at this pressure. An increase in pressure (at 33 K) results in a broadening of some of the diffraction peaks (Fig. 4(a)) above 29 GPa. In order to account the broadening , the diffraction data above 29 GPa were refined using an orthorhombic phase (space group *Fmmm*), tetragonal (*I4/mmm*) and a combination of tetragonal and orthorhombic phases (Fig. 4(b)) respectively. Here we show an example (Fig. 4(b)) of such an analysis for a measurment at 34.1 GPa and 33 K. The analysis of the diffraction data clearly indicate the presence of an additional tetragonal phase above 29 GPa. The coexistence of tetragonal and orthorhombic phases (Fig. 4) suggests onset of the first order phase transition at 29 GPa (at 33 K). The ratio of orthorhombic and tetragonal phases at 29 GPa is 74 % and 26 % respectively. The percentage of the orthorhombic phase was found to decrease on increasing the pressure up to 34.1 GPa (shown later in Fig. 6). During heating, the diffraction profiles were collected at 100 K (at 34.1 GPa), 200 K (at 34.9 GPa) and 300 K (36.7 GPa). The diffraction pattern at 300 K and 36.7 GPa could be indexed using only the collapsed tetragonal phase. Finally, the diffraction profiles were collected while lowering the pressure at 300 K. We noticed that below 20 GPa at 300 K the collapsed phase reverted back to the ambient tetragonal phase.

The effects of pressure inhomogineity on the phase transition behaviour of FeAs compounds have been discussed in the literature [21, 22]. Recently, pressure dependent electrical resistivity of $BaFe_2As_2$ has been measured up to 16 GPa [21]. The liquid Fluorinert was used as pressure transmitting medium. Above 1.2 GPa, the solidification of Fluorinert may yield moderate inhomogeneous pressure distributions. Duncan et al [22] has used three different pressure transmitting media e.g. pentane–isopentane, Daphne oil and steatite for measurements of electrical resistivity of $BaFe_2As_2$ up to about 5 GPa. All these pressure media have their own intrinsic level of hydrostaticity and yield moderate inhomogeneous pressure distributions at very low pressures. The authors [22] find that the pressure–temperature phase diagram of $BaFe_2As_2$ is extremely sensitive to the pressure-transmitting medium used for the experiment and, in particular, to the level of resulting uniaxial stress. An increasing uniaxial pressure component in this system quickly reduces the spin density wave order and favours the appearance of superconductivity.

In the present measurements, we have used helium as a pressure-transmitting medium, which give the best hydrostatic conditions [36]. During the measurements, we have determined the pressure using two ruby balls. The pressure difference determined from ruby balls was always below 0.1-0.2 GPa. Non-hydrostatic effects might not have influence on the results obtained from our studies on these compounds.

Figures 5 and 6 show pressure dependence of the structural parameters (lattice parameters, volume and $c_t/a_t$ obtained from Rietveld refiements for BaFe$_2$As$_2$ at 300 K (tetrgonal phase) and 33 K (in orthorhombic phase) in pressure increasing and decreasing cycles respectively. For easy comparison, orthorhombic lattice parameters ($a$, $b$, and $c$) are converted into the equivalent tetragonal lattice parameters ($a_{ot}$, $b_{ot}$ and $c_{ot}$) using the relation $a_{ot} = a/\sqrt{2}$, $b_{ot}=b/\sqrt{2}$ and $c_{ot}=c$. It is clear from figure 5 that upon increasing pressure at 300 K, the $a$ lattice parameter first decreases (up to 22 GPa) and then increases (up to 32 GPa) and again further decraeses up to 56 GPa. The $c$ lattice parameter decreases with pressure in the entire range of our measurements.

We have fitted the pressure volume data using Birch-Murnaghan equation of state separately in the pressure range of 0-20 GPa (tetragonal phase) and 32-56 GPa (collapsed tetragonal phase) at 300 K. It can be clearly seen from Fig. 5 that the volume decreases by about 1.4 % across the phase transition. Using a similar fitting procedure, at the transition, the $a_t$ lattice parameter was found to increase (1.75 %) with compression while $c_t$ decreased (4.9 %). The variation of $a_t$, $c_t$, $c_t/a_t$ and $V$ at 300 K in the pressure increasing and decreasing cycles (Fig. 5) clearly indicate hysteresis during the phase transition, and confirm the first order nature of phase transition. The critical value of pressure for the phase transition is 26 GPa as obtained from the mid-point of the hystresis loop. Similar observations have already been seen for CaFe$_2$As$_2$ [12, 13]. The variation of volume and $c_t/a_t$ as a function of $P/P_c$ ($P$ and $P_c$ are the values of the applied and phase transition pressure respectively) in both the BaFe$_2$As$_2$ and CaFe$_2$As$_2$ (Fig. 7) appears to be nearly identical. X-ray diffraction measurements for CaFe$_2$As$_2$ show [12, 13] that during the structural phase transition from the tetragonal to the collapsed phase, the $c_t$ lattice parameter contracts by about 9%, whereas $a_t$ expand by about 1.5%. Furthermore, in CaFe$_2$As$_2$ there is a continuous evolution of $c_t/a_t$ as a function of pressure. The authors of Ref. [12, 13] have not reported any phase co-existence during the transition from the tetragonal to the collapsed tetragonal phase. However, there is a small hysteresis in the $a_t$, $c_t$ and $c_t/a_t$ values. Similar behavior has been found in our measurements (Fig. 5) for BaFe$_2$As$_2$. For comparison, the results from a molecular dynamics (MD) simulation [26] of a pressure induced phase transition in these compounds are also shown in Fig. 5. We find that transition behaviour for CaFe$_2$As$_2$ as obtained from MD simulation is in agreement with the experimental data. However, for BaFe$_2$As$_2$, the calculation does not agree with the experimental observations.

Similar to CaFe$_2$As$_2$ [12, 13], observation of a hysteresis in the variation of lattice parameters in the pressure increasing and decreasing cycles of BaFe$_2$As$_2$ at 300 K, across the tetragonal to collapsed tetragonal phase transition, suggest first order nature of phase transition. On the other hand, at 33 K the lattice parameters, volume and $2c_{ot}/(a_{ot}+b_{ot})$ (equivalent tetragonal values) are found to decrease (Fig. 6) with increasing pressure. Above 29 GPa, we found a coexistence of orthorhombic and tetragonal phases. It is interesting to note that the transition to the collapsed phase in BaFe$_2$As$_2$ occurs at nearly the same volume of about 165 Å$^3$ on compression at 300 K or 33 K.

The data for BaFe$_2$As$_2$ have been measurements in both the pressure increasing and decreasing cycles at 300 K. The critical value of pressure for the phase transition is 26 GPa as obtained from the mid-point of the hysteresis loop. However, at 33 K measurements are carried out only in pressure increasing cycle. The coexistence of tetragonal and orthorhombic phases (Fig. 4) suggests onset of the phase transition at 29 GPa (at 33 K). The phase transition behavior in BaFe$_2$As$_2$ seems to be different in comparison with CaFe$_2$As$_2$. The critical pressure for the phase transition for BaFe$_2$As$_2$ at 33 K and 300 K is around 29 GPa and 26 GPa respectively. However, this is in contrast to the reported behavior [12, 13] in CaFe$_2$As$_2$ where the phase transition at lower temperature (50 K) occurs at a lower pressure (0.3 GPa) in comparison to 1.7 GPa at 300 K.

The electronic properties of the FeAs superconductors are sensitively controlled by distortions of the FeAs$_4$ tetrahedra in terms of As-Fe-As bond angle and Fe-As bond length. Pressure dependence of the As-Fe-As bond angle and As-Fe bond length of BaFe$_2$As$_2$ at 300 K (tetrgonal phase) and 33 K (in the orthorhombic phase) are shown in Figure 8. It is clear from this figure that at 300 K (in the tetragonal phase), the two As-Fe-As bond angles were close to the ideal tetrahedral value of 109.47$^{\circ}$. However, with an increase of pressure the deviation from the ideal tetrahedral angle increases. There is an anomalous increase in the difference between the two As-Fe-As bond angles starting at about 10 GPa and 22 GPa (Fig. 8(a)). The latter anomaly can be associated with structural transition to the collapsed phase. However lattice parameters do not indicate any appreciable change at about 10 GPa. The anomaly in the As-Fe-As bond angles at 10 GPa could have an electronic origin.

Recent high pressure resistivity measurements indicate suppression of the magnetic transition in BaFe$_2$As$_2$ [21] at 10.5 GPa. The anomaly in the As-Fe-As bond angles at about 10 GPa and 33 K (Fig.

8(b)) may be due to suppression of antiferromagnetic ordering. The tetragonal transition at 29 GPa (at 33 K), as observed in our experiments, is far above 10.5 GPa. Earlier it has been shown that the structural and magnetic transtions CaFeAsF and SrFeAsF compounds [6, 24] do not occur at the same temperature. It appears that the magnetic and structural transitions occur at 10 and 29 GPa (at 33 K) respectively. The inelastic neutron scattering measurements [23] as well as ab-initio calculations [20] show that spin fluctuations in FeAs compounds are present at all temperatures up to 300 K at ambient pressure. We believe that the anomaly in the As-Fe-As bond angles at 10 GPa (Fig. 8(a)) may be due to suppression of paramagnetic spin fluctuations at 300 K.

Ab-initio calculations [20] also show that the strong interaction between As ions in FeAs compounds is controlled by the Fe-spin state. Reducing the Fe-magnetic moment weakens the Fe-As bonding and, in turn, increases the As-As interactions. The loss of Fe-moment at 10.5 GPa in $BaFe_2As_2$ [22] may increase the As-As interaction along the c-axis. This, in turn, would change the $z$ parameter of the As atom and may lead to an increase in the distortion of the As-Fe-As bond angle at 10.5 GPa in our measurements at 34 K and 300 K.

It can be seen that in both phases at 300 K the compressibility along the $a$-axis is smaller than that along the $c$-axis. The pressure-volume data were fitted by a third order Birch-Murnaghan equation of state in order to determine the bulk modulus $B$ at zero pressure and its pressure derivative $B'$. The obtained parameters are $B = 65.7 \pm 0.8$ GPa, $B' = 3.9 \pm 0.1$ for the tetragonal phase (0- 20 GPa) and $B = 153 \pm 3$ GPa, $B' = 1.8 \pm 0.1$ for the collapsed tetragonal phase (from fitting of data from 32- 56 GPa) at 300 K. The $B$ and $B'$ values extracted from the pressure volume relation in the orthothombic phase (33 K, from fitting of data from 1-34 GPa) are $82.9 \pm 1.4$ GPa and $3.4 \pm 0.1$ respectively. The fitted ambient pressure volumes for the tetragonal and collapsed tetragonal phase at 300 K are $V_o = 204.3 \pm 0.1$ Å$^3$ and $181.6 \pm 0.7$ Å$^3$, respectively. However, $V_o$ for the orthorhombic phase at 33 K is $201.78 \pm 0.13$ Å$^3$.

The bulk modulus values are very close to that obtained from the high pressure measurements [18] of $LaFeAsO_{0.9}F_{0.1}$ (B=78 GPa). The $c_t/a_t$ ratio for $BaFe_2As_2$ varies from 3.3 to 2.55 upon an increase of pressure to 56 GPa. It should be noted that $c_t/a_t$ of the Ba compound reaches a value of 2.92 before the transition to the collapsed phase, which is nearly the same as for $CaFe_2As_2$ at ambient pressure. However, in the case of $CaFe_2As_2$, the transition to the collapsed phase is at very low pressures of 1.7 GPa at 300 K. The FeAs$_4$ tetrahedral volumes in the tetragonal phase at ambient conditions and in the orthorhombic phase at 33 K and 1 GPa are about 22 Å$^3$. However, the collapsed transition at 300 K as well as 33 K starts when the FeAs$_4$ tetrahedra are compressed (Fig. 9) below about 17.5 Å$^3$. We also found that the FeAs$_4$ tetrahedra are much less compressible (Fig. 9) and are much more distorted in the collapsed phase compared to their nearly regular shape in the ambient pressure tetragonal phase.

**B. High pressure phase stability of CaFe$_2$As$_2$**

It is well established in the literature that at ambient pressure, CaFe$_2$As$_2$ undergoes [7] a first order transition from a high temperature nonmagnetic tetragonal phase to an antiferromagnetic orthorhombic phase at $T$ = 172 K. The application of modest pressures ($P$>0.23 GPa) at low temperatures of about 40 K, transforms the antiferromagnetic orthorhombic phase to a different non-magnetically ordered collapsed tetragonal structure. The lattice parameters are found to change significantly as a function of pressure, with a dramatic decrease in both the unit-cell volume and the $c_t/a_t$ ratio.

To explore the possibilities of a structural phase transition from collapsed tetragonal to another phase with pressure, we carried out powder synchrotron diffraction experiments at high pressures. The sample was first cooled to 40 K at about 0.5 GPa. Once the temperature was stabilized at 40K, the pressure was increased up to ~ 34 GPa. The diffraction data (Fig. 10) does not show any appreciable change in all the studied pressure range, i.e. CaFe$_2$As$_2$ remained in the collapsed tetragonal phase.

We also performed another experiment where we collected powder diffraction data at 300 K up to 51 GPa. The sample undergoes a structural phase transition (Fig. 10) into the collapsed phase at about 2 GPa. This observation is in agreement with that already reported in the literature [11-13]. The neutron and x-ray diffraction experiments [11-13] carried out up to 5 GPa show a similar transition from tetragonal to the collapsed tetragonal phase at 1.7 GPa. As shown in Fig. 9, all the peaks in the diffraction profiles are well accounted, using a collapsed tetragonal phase at the highest pressure. Further, there is no signature of any post collapsed phase transition in CaFe$_2$As$_2$ up to 51 GPa (~34 GPa) at 300 K (40 K).

Figures 11 and 12 show the variation of lattice parameters, $c_t/a_t$, volume, As-Fe-As bond angle and Fe-As bond length obtained after the Rietveld analysis of powder synchrotron diffraction data as a function of pressure in compression cycles (Fig. 1) of CaFe$_2$As$_2$ at 40 K and 300 K. It is clear from Fig. 11 that the structural parameters ($a$, $c$, $c_t/a_t$, $V$) decrease with increasing pressure. We find that at 0.5

GPa and 40 K, the As-Fe-As bond angles show a small deviation from the ideal value of 109.47° for an ideal FeAs$_4$ tetrahedron. A further increase of pressure up to 4 GPa shows an increase in deviation from an ideal tetrahedron angle indicating an increasing distortion of the FeAs$_4$ tetrahedron in the collapsed phase. The deviation is found to decrease with a further increase of pressure up to 10 GPa and then it remains almost constant up to the highest measured pressure of 37 GPa at 40 K. At 300 K, we find that in the parent tetragonal phase at 0.5 GPa and 300 K the deviation of the As-Fe-As angles (Fig. 12) is smaller than the deviation at a similar pressure at 40 K. As expected, the deviation was found to increase with an increase of pressure to 2 GPa during the tetragonal to collapsed tetragonal phase transition. We also find that as pressure is increased to 51 GPa at 300 K, the As-Fe-As bond angle data show anomalies at about 20 GPa and 40 GPa and may be linked with spin fluctuation. However, resistivity data of CaFe$_2$As$_2$ are not available at pressures above 1.8 GPa [31]. Such data would be useful for correlating the distortions in the bond angle data. As explained earlier for BaFe$_2$As$_2$, the increases in distortion of the bond angles are related to the reduction in the Fe-magnetic moment with increase of pressure. High-pressure experiments on FeSe at ambient temperature also show [19] a similar behaviour where such an increase in distortion of the Se-Fe-Se bond angles of FeSe$_4$ was found at very low pressures of 1 GPa.

It is also interesting to notice that the variation of the Fe-As bond length in the collapsed phase at 300 K shows a decrease of about 10 % (Fig. 12) upon an increase of pressure from 2 GPa to 51 GPa. The pressure variation of FeAs$_4$ tetrahedral volume in CaFe$_2$As$_2$ is shown in Fig. 9. The collapsed transition in CaFe$_2$As$_2$ starts when the FeAs$_4$ tetrahedral volume (at 300 K as well as at 40 K) reaches a value of about 20.5 Å$^3$. It is clear from Fig. 7 that in the collapsed phase the FeAs$_4$ tetrahedra are much less compressible at 40 K than at 300 K. This is also consistent with the pressure variation of As-Fe-As bond angle and Fe-As bond length (Fig. 12) in CaFe$_2$As$_2$ at 300 K and 40 K. We recall, as shown earlier, the collapsed transition in BaFe$_2$As$_2$ starts as the FeAs$_4$ tetrahedral volume approaches about 17.5 Å$^3$. We find significantly different pressure dependences of the polyhedral volume for the two temperatures (300 K and 40 K) above 12 GPa. Since the transformation pressure in BaFe$_2$As$_2$ is much higher than that in CaFe$_2$As$_2$, the Ba-compound might show significant temperature dependence if measurements are extended up to much higher pressures in comparison to the transition pressure.

The calculated variation of volume with pressure for CaFe$_2$As$_2$ using ab-initio methods [20] gives bulk modulus values of 56.2 GPa and 81.6 GPa for the tetragonal and collapsed tetragonal state, respectively. However, from our measurements the $B$ and $B'$ values at 300 K in the collapsed phase (from fitting of data from 4.5- 56 GPa) are 74.8 ± 1.2 GPa and 4.8 ± 0.1, respectively, while at 40 K in the collapsed phase these values (from fitting of data from 4- 37.8 GPa) found to be 80.2 ± 3.4 GPa and 5.4 ± 0.2, respectively. The fitted ambient pressure volumes for the collapsed tetragonal phase at 300 K and 40 K are 170.1 ± 0.2 Å$^3$ and 167.7 ± 0.4 Å$^3$, respectively. The comparison of pressure variation of $c_t/a_t$ in both the compounds ($CaFe_2As_2$ and $BaFe_2As_2$) shows (Fig. 7) that the structure of FeAs compounds, becomes unstable as $c_t/a_t$ approaches a value of 2.9, completely transforms to the collapsed phase (Fig. 7) near the $c_t/a_t$ value of 2.7. Furthermore, the collapsed phase transition in $CaFe_2As_2$ occurs (Fig. 7) at nearly the same volume of about 165 Å$^3$ upon compression as in the case of $BaFe_2As_2$ at both 300 K and 33 K.

## IV. Conclusions

In conclusion, we have carried out high-pressure powder synchrotron x-ray diffraction studies over a wide range of pressures and temperatures. Detailed analysis of the data for $BaFe_2As_2$ at 300 K, show a phase transition from the tetragonal to the collapsed tetragonal phase at about 26 GPa that remains stable up to 56 GPa. On the other hand, compression of the sample at 33 K, reveals onset of a transformation from orthorhombic to tetragonal phase at 29 GPa. Measurements on $CaFe_2As_2$ confirm a transition to a collapsed phase as reported in the literature. We have not found any evidence of further post collapsed tetragonal phase transition in $CaFe_2As_2$ up to 51 GPa (at 300 K) and 37.8 GPa (at 40 K). The transition to a collapsed phase occurs in the two compounds at nearly the same values of unit cell volume and $c_t/a_t$ ratio. High pressure resistivity measurements in $EuFe_2As_2$ show [16] that superconducting transition temperature jumped from 22 K to 41 K at the collapsed phase transition. It would be interesting to search for superconductivity in the high pressure phase of $BaFe_2As_2$.

**Acknowledgments**

R. Mittal and S. K. Mishra thank Department of Science and Technology (DST), India for providing financial support to carry out synchrotron x-ray diffraction at European Synchrotron Radiation Facility, Grenoble, France. We thank M. Hanfland and W. Crichton for loading the helium into the cells for our measurements carried out at ESRF.

**TABLE I**. Refined results of the crystal structure for $BaFe_2As_2$ at selected pressure and at temperatures of 300 and 33 K. Atomic positions for space group *I4/mmm*: Ba (2a) (0 0 0); Fe(4d) (½ 0 ¼ ) and As (4e) (0 0 z). Atomic positions for space group *Fmmm* : Ba(4a)(0 0 0), Fe (8f) ( ¼, ¼, ¼ ), As(8i) (0 0 z). The measurements were carried out using a focused x-ray monochromatic beam of wavelength=0.3738 Å. The data collected up to 20º has been used to determine the reported parameters.

| Temperature (K) | **300** | **300** | **33** | **33** |
|---|---|---|---|---|
| Pressure (GPa) | **0.2** | **56** | **1.3** | **34.1** |
| Space group | *I 4/ m m m* | *Fmmm* | *Fmmm* | *Fmmm+I4/mmm* |
| a (Å) | 3.9590(1) | 3.7796(2) | 5.5447(3) | 5.2571(8)/3.7643(6) |
| b (Å) | 3.9590(1) | 3.7796(2) | 5.5773(3) | 5.2738(9)/3.7643(6) |
| c (Å) | 13.005(3) | 9.593(2) | 12.852(6) | 11.311(3)/10.828(3) |
| V(Å) | 203.85(1) | 137.07(1) | 397.45(4) | 313.70(7)/153.43(5) |
| z | 0.3560(2) | 0.3662(4) | 0.3575(4) | 0.3657(8)/0.3676(8) |
| Rp | 9.82 | 10.7 | 17.6 | 19.1 |
| Rwp | 14.1 | 14.8 | 22.7 | 25.2 |
| Rexp | 8.30 | 9.60 | 11.92 | 14.65 |
| $\chi^2$ | 2.88 | 2.37 | 3.03 | 2.96 |
| No. of reflections | 78 | 56 | 66 | 60 |

**TABLE II.** Refined results of the crystal structure for $CaFe_2As_2$ at selected pressure and at temperatures of 300 and 40 K. Atomic positions for space group *I4/mmm*: Ca (2a)(0 0 0); Fe(4d) (½ 0 ¼ ) and As (4e) (0 0 z). The measurements were carried out using a focused x-ray monochromatic beam of wavelength=0.3738 Å. The data collected up to 20° has been used to determine the reported parameters.

| Temperature (K) | **300** | **300** | **40** | **40** |
|---|---|---|---|---|
| Pressure (GPa) | **1** | **50.8** | **4.6** | **37.8** |
| **Space group** | *I 4/ m m m* | *I 4/ m m m* | *I 4/ m m m* | *I 4/ m m m* |
| a=b(Å) | 3.9052(1) | 3.6727(3) | 3.9396(1) | 3.7439(3) |
| c (Å) | 11.402(2) | 9.3173(5) | 10.258(4) | 9.4616(8) |
| V(Å) | 173.88(2) | 125.68(2) | 159.22(3) | 132.62(2) |
| Z | 0.3726(4) | 0.3843(7) | 0.3690(2) | 0.3726(3) |
| Rp | 18.0 | 15.4 | 17.8 | 18.2 |
| Rwp | 22.5 | 20.6 | 22.3 | 24.1 |
| Rexp | 12.62 | 12.85 | 12.95 | 14.77 |
| $\chi^2$ | 3.18 | 2.58 | 2.95 | 2.65 |
| No. of reflections | 62 | 44 | 52 | 37 |

FIG. 1 (Color online) The pressure-temperature conditions for measurement of $BaFe_2As_2$ and $CaFe_2As_2$. The circles, squares and stars represent experimental data points in different cycles of our measurements. The solid and open circles correspond to the data points for $BaFe_2As_2$ in the second cycle of measurements while increasing and decreasing pressure, respectively. The solid lines through the symbols are guides to the eye. Arrows indicate the sequence of measurements during the experiment.

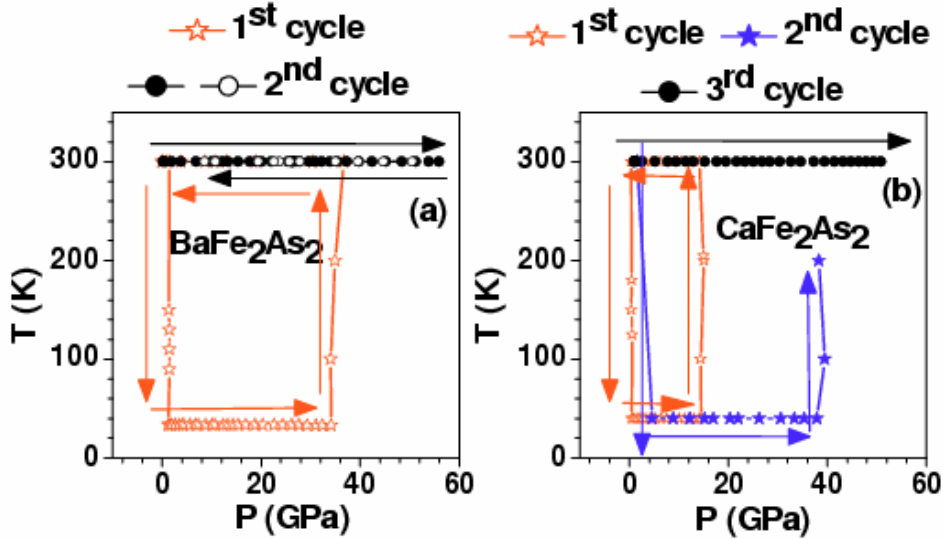

FIG. 2 (Color online) (a) Evolution of the powder synchrotron X-ray diffraction patterns of $BaFe_2As_2$ at 300 K. (b) Observed (solid circle: black), calculated (continuous red line), and difference (bottom blue line) profiles obtained after the Rietveld refinement of X-ray diffraction patterns of $BaFe_2As_2$ using tetragonal space group (*I4/mmm*) at selected pressures at 300 K.

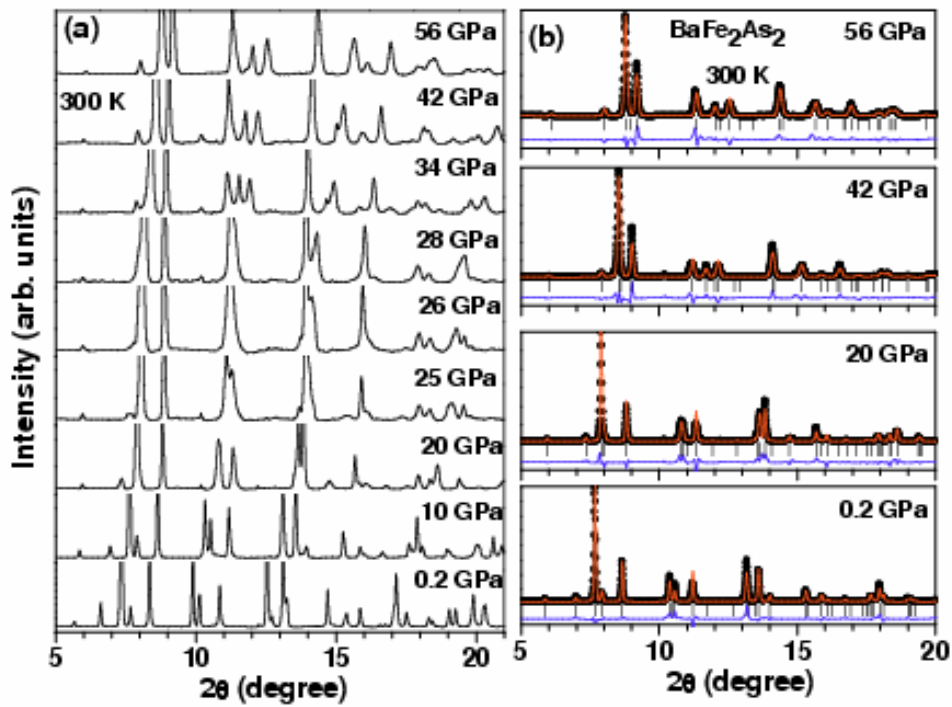

FIG. 3 (Color online) Tetragonal to orthorhombic phase transition in BaFe$_2$As$_2$ at 1.3 GPa and 130 K. The refinement of the diffraction pattern with an (a) orthorohmbic phase (space group *Fmmm*) and (b) tetragonal (*I4/mmm*) phase. The fitted profile clearly indicates an orthorhombic phase at 1.3 GPa and 130 K.

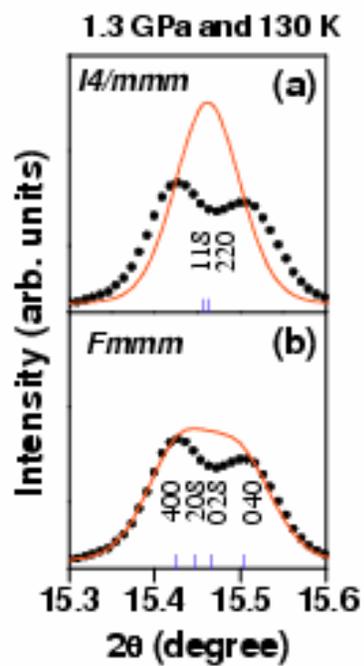

FIG. 4 (Color online) (a) Observed (solid circle: black), calculated (continuous red line), and difference (bottom blue line) profiles obtained after the Rietveld refinement of $BaFe_2As_2$. The diffraction profiles at 1.3 GPa and 22 GPa are refined using an orthorohmbic phase (space group *Fmmm*), while the profiles at 29 GPa and 34.1 GPa are refined using a combination of tetragonal (*I4/mmm*) and orthorohmbic (space group *Fmmm*) phases. Upper and lower vertical tick marks above the difference profiles at 29 GPa and 34.1 GPa indicate peak positions of orthorhombic (*Fmmm*) and tetragonal (*I4/mmm*) phases, respectively. (b) The refinement of the diffraction pattern at 34.1 GPa and 33 K with an orthorhombic phase (space group *Fmmm*), a tetragonal (*I4/mmm*) phase, and a combination of tetragonal (*I4/mmm*) and orthorohmbic (space group *Fmmm*) phases.

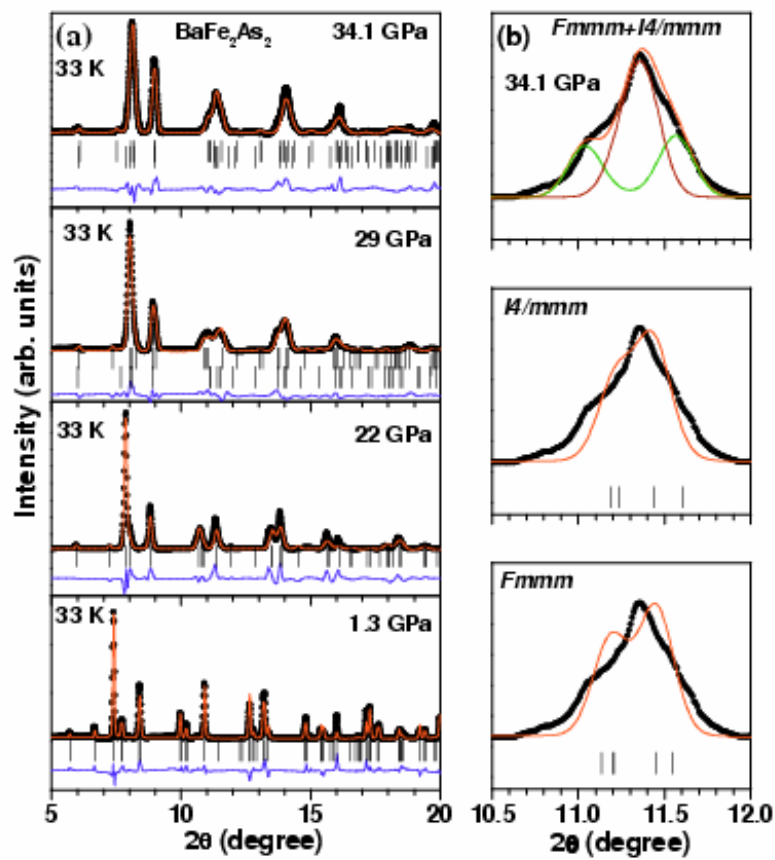

FIG. 5 (Color online) Pressure dependence of the structural parameters (lattice parameters, volume) and $c_t/a_t$ of $BaFe_2As_2$ at 300 K (tetragonal phase) in pressure increasing (P up) and decreasing (P down) cycles. The solid and open symbols correspond to the measurements in pressure increasing and decreasing cycles. The right hand side figure in the lower row shows fitting of pressure-volume data (pressure increasing cycle) at 300 K to the third order Birch Murnaghan equation of state in the tetragonal (T) and collapsed tetragonal (cT) phase. The critical value of pressure ($P_c$) for the phase transition at 300 K is 26 GPa. The critical value of pressure represents the mid-point of the hysteresis loop.

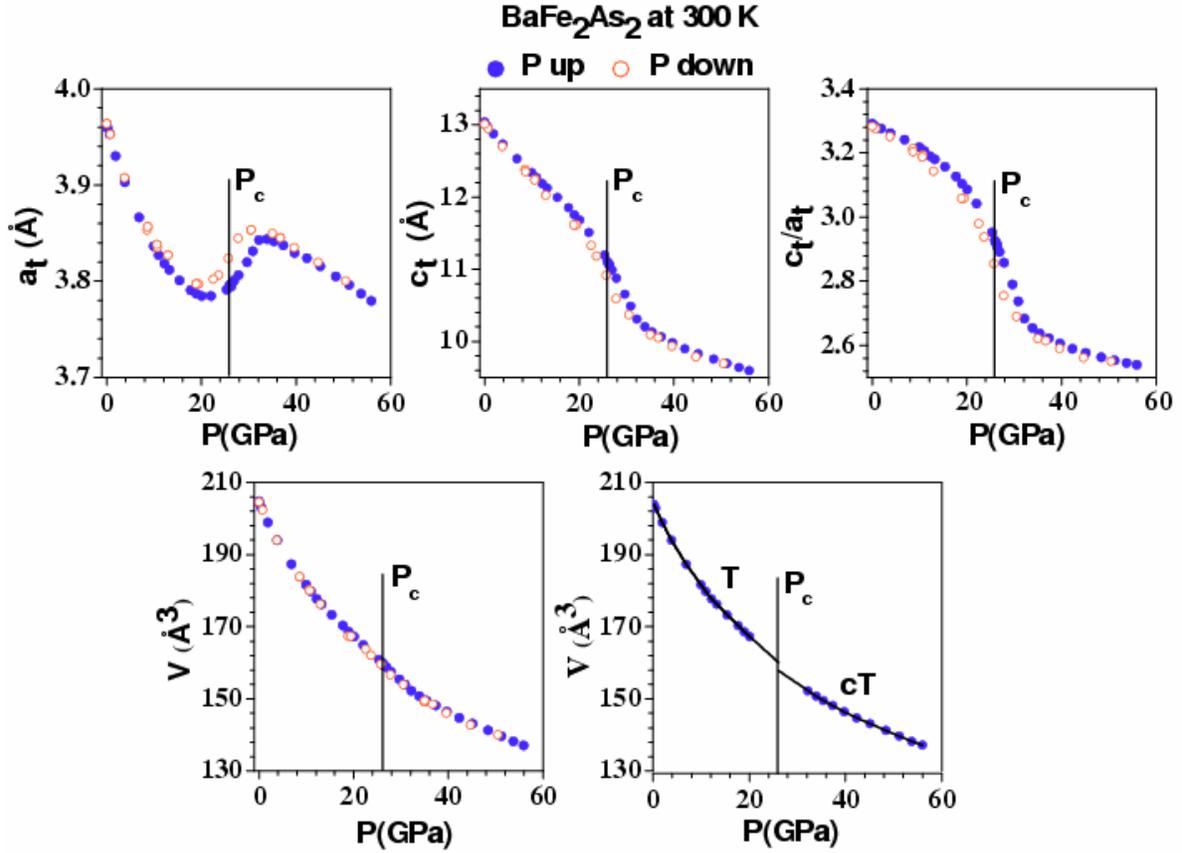

FIG. 6 (Color online) Pressure dependence of the structural parameters (lattice parameters, volume) and $c_t/a_t$ of BaFe$_2$As$_2$ at 33 K (in orthorohmbic phase) in pressure increasing (P up) cycle. For easy comparison, orthorhombic lattice parameters ($a$, $b$, and $c$) are converted into the equivalent tetragonal lattice parameters ($a_{ot}$, $b_{ot}$ and $c_{ot}$) using the relation $a_{ot} = a/\sqrt{2}$, $b_{ot}=b/\sqrt{2}$ and $c_{ot}=c$. The open symbols above 29 GPa correspond to the structural data ($a_t$, $c_t$ and $c_t/a_t$) in the collapsed tetragonal phase. The percentage variation of orthorhombic phase during our measurements is shown in the inset of the lower row. The critical value of pressure ($P_c$) for the phase transition at 33 K is 29 GPa.

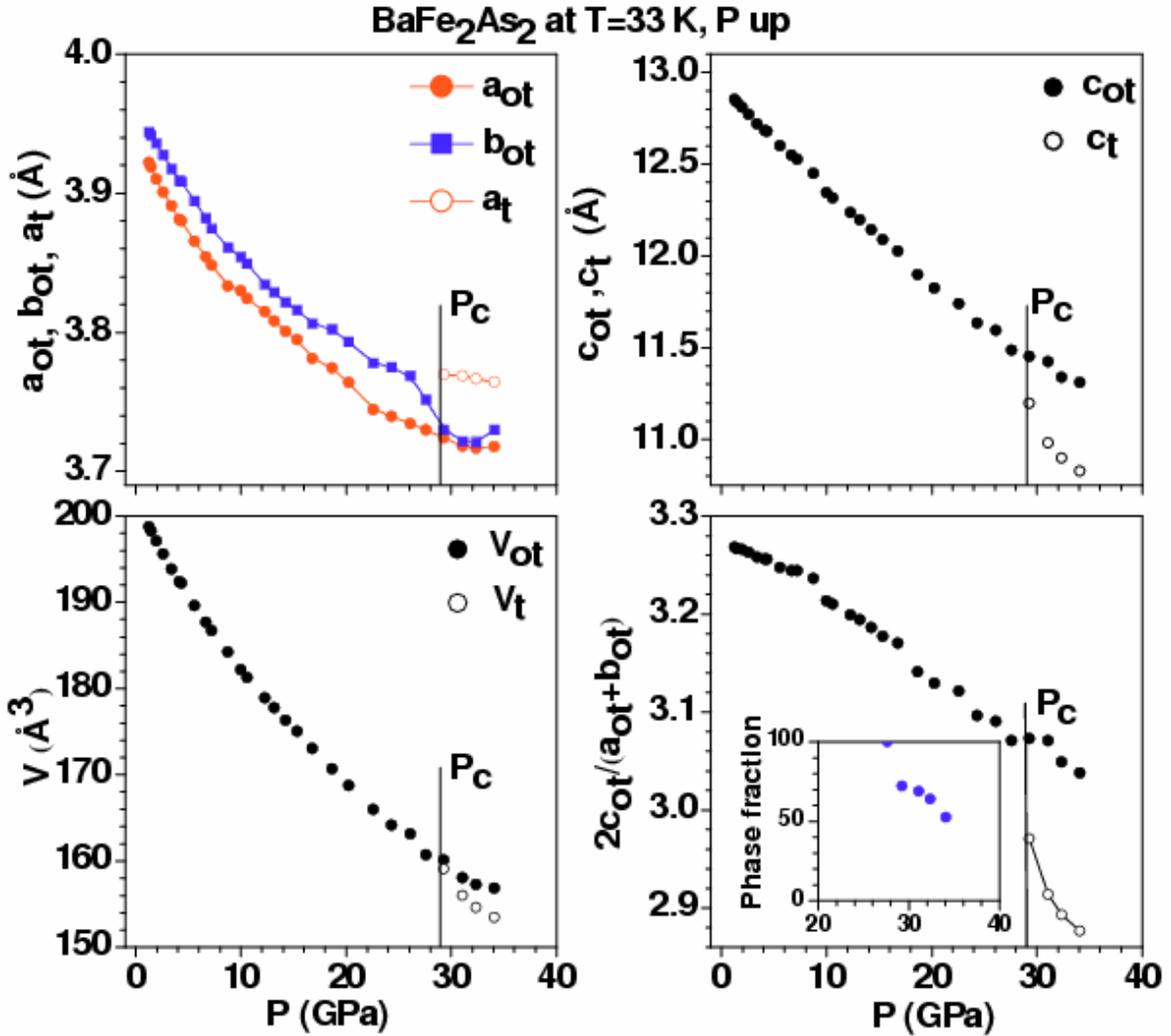

FIG. 7 Volume and $c_t/a_t$ as a function of $P/P_c$ at 300 K. $P$ and $P_c$ are the values of the applied and phase transition pressure, respectively. The $P_c$ values are 26 GPa and 1.7 GPa for $BaFe_2As_2$ and $CaFe_2As_2$, respectively. The comparsion between the experimental data and molecular dynamics simulations is shown in middle and last row. The solid squares and solid circles are the data from our experiment during pressure increasing cycles at 300 K. For comparison, the experimental data published by Goldman et al [13] for $CaFe_2As_2$ are shown by open circles. The solid and open triangles in the middle and last row correspond to the results obtained from molecular dynamics (MD) simulation [26] for $BaFe_2As_2$ and $CaFe_2As_2$, respectively. The solid lines through the symbols are guides to the eye.

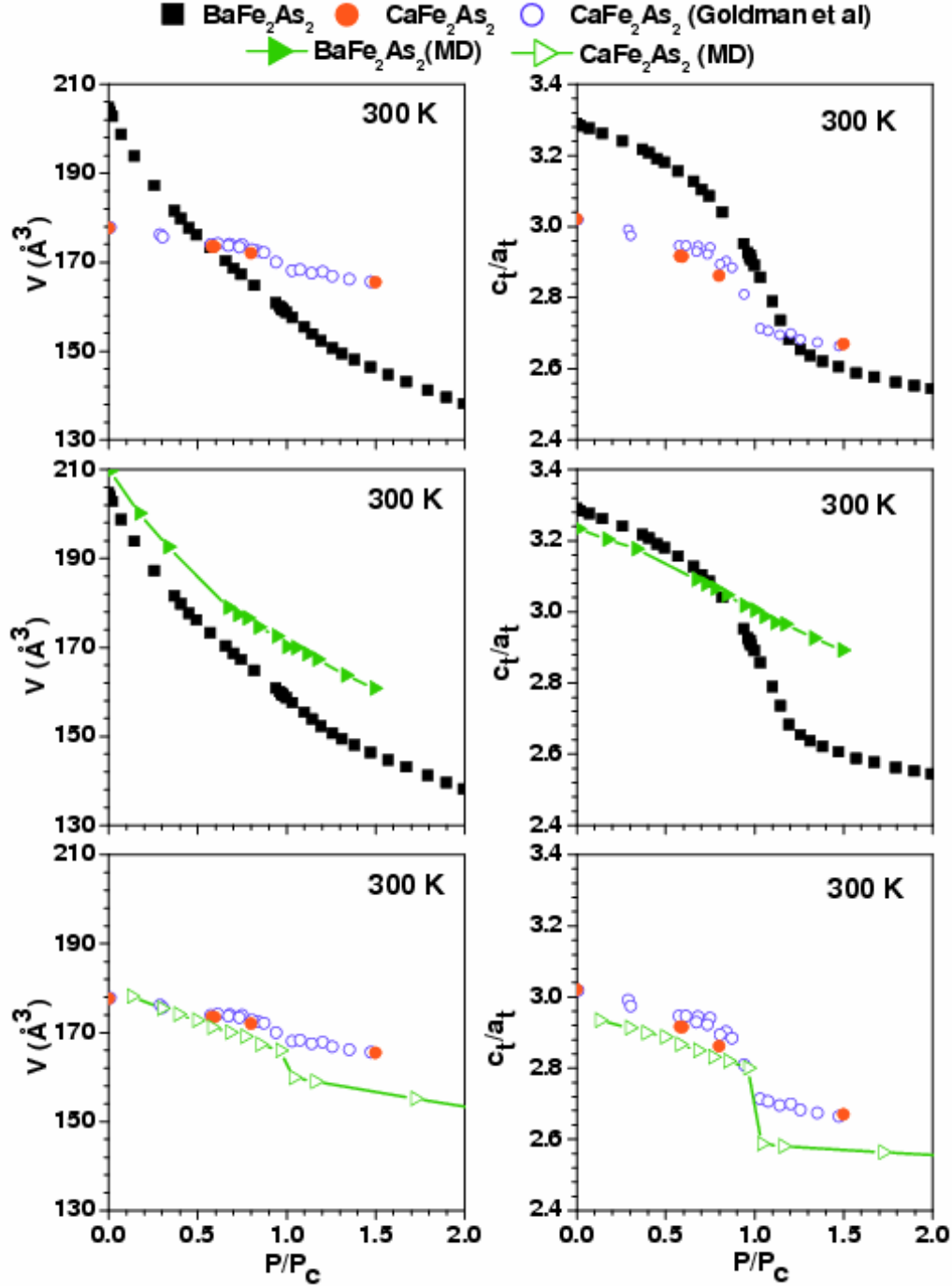

FIG. 8 (Color online) Pressure dependence of the As-Fe-As Bond angle and As-Fe bond length of $BaFe_2As_2$ at (a) 300 K (tetragonal phase) and (b) 33 K (in orthorohmbic phase). The solid lines through the symbols are guides to the eye. The error bars at 300 K are comparable to the size of the symbol.

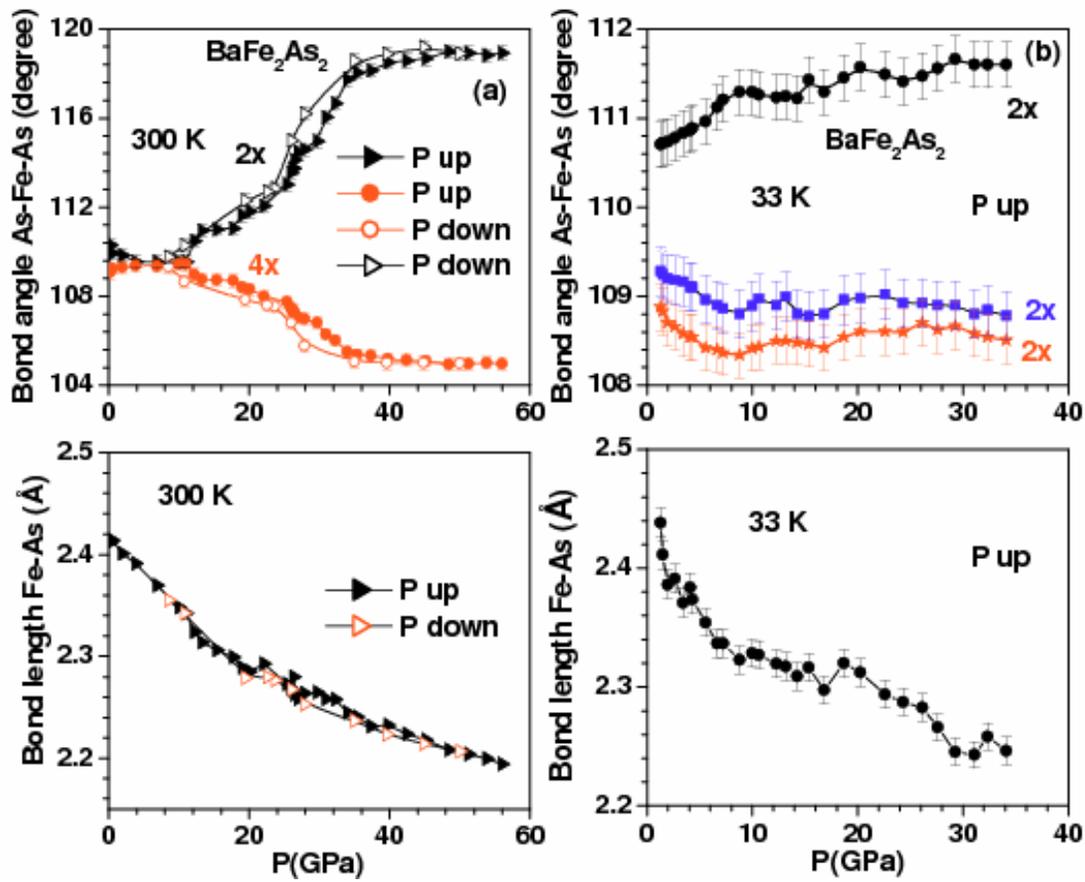

FIG. 9 (Color online) Pressure variation of $FeAs_4$ tetrahedral volume in $BaFe_2As_2$ and $CaFe_2As_2$.

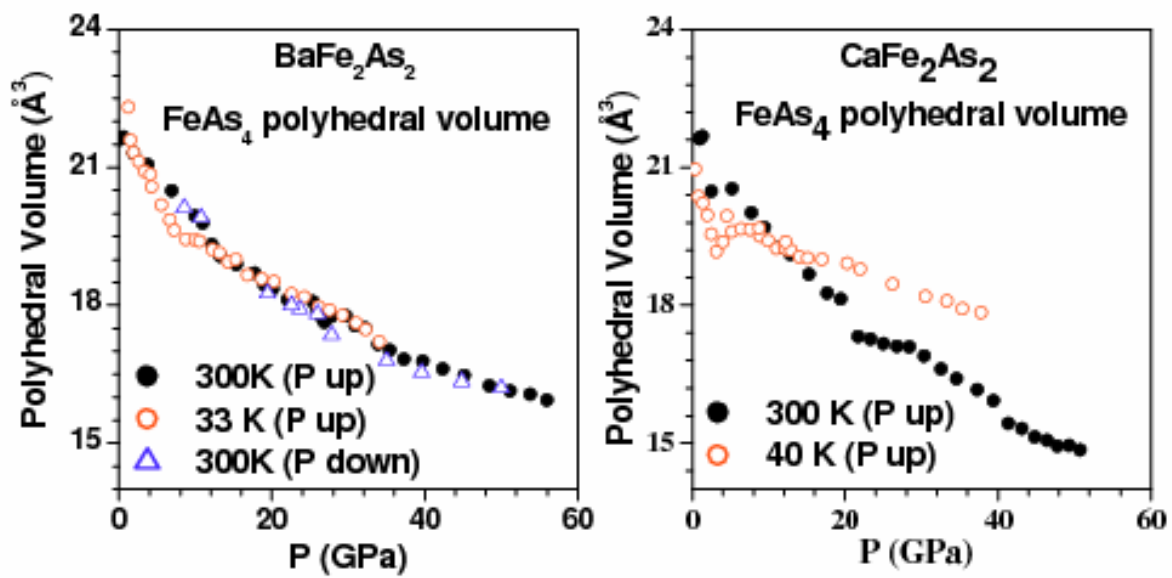

FIG. 10 (Color online) Observed (open circle), calculated (continuous line), and difference (bottom line) profiles obtained after the Rietveld refinement of CaFe$_2$As$_2$ at selected pressures and 40 K and 300 K. The diffraction profiles are refined using the tetragonal space group *I4/mmm*.

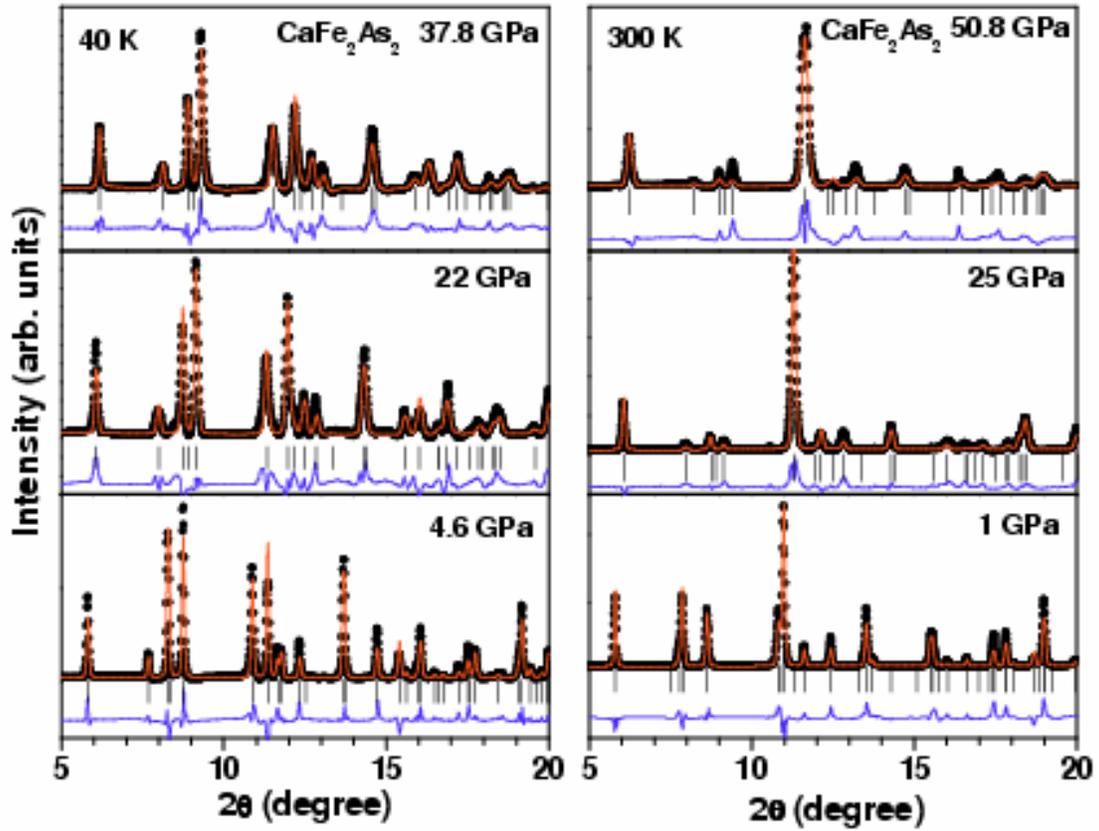

FIG. 11 (Color online) Pressure dependence of the structural parameters (lattice parameters, volume) and $c_t/a_t$ of CaFe$_2$As$_2$ at (a) 300 K and (b) 40 K in pressure increasing (P up) and decreasing (P down) cycles. Solid and open symbols correspond to the measurements in pressure increasing and decreasing cycles.

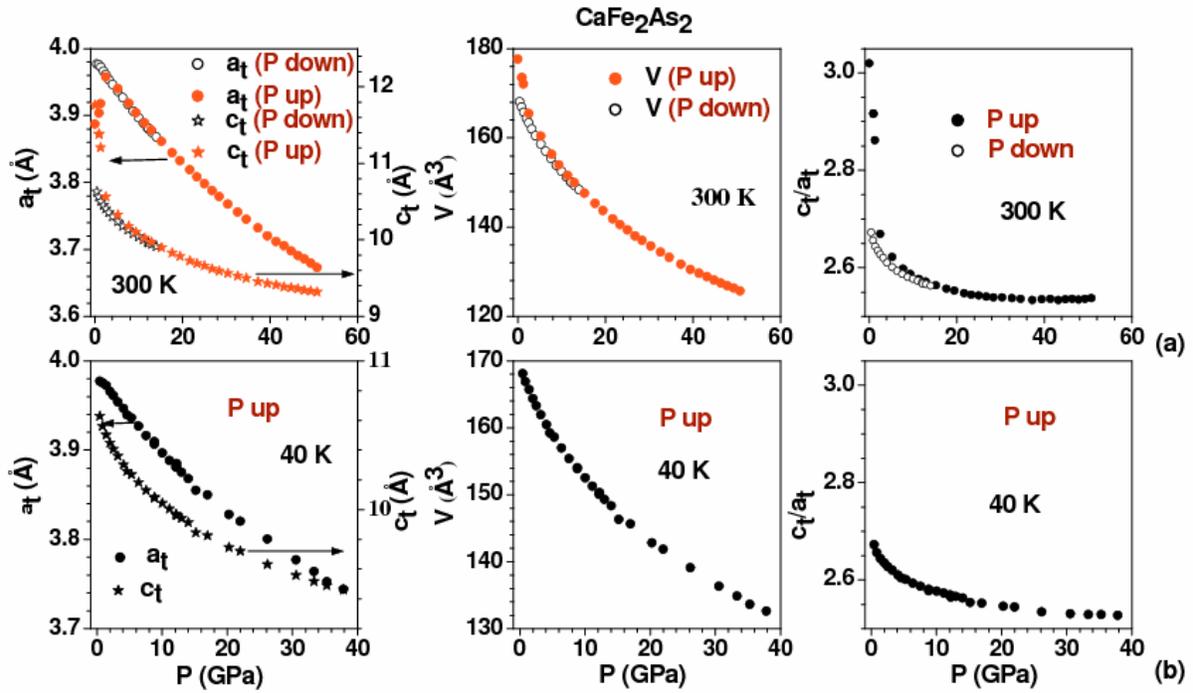

FIG. 12 (Color online) Pressure dependence of the As-Fe-As Bond angle and As-Fe bond lentgh of CaFe$_2$As$_2$ at (a) 300 K and (b) 33 K, respectively. The bond angles are plotted only for data taken during pressure increasing cycles at 300 K and 33 K. The solid lines through the symbols are guides to the eye.

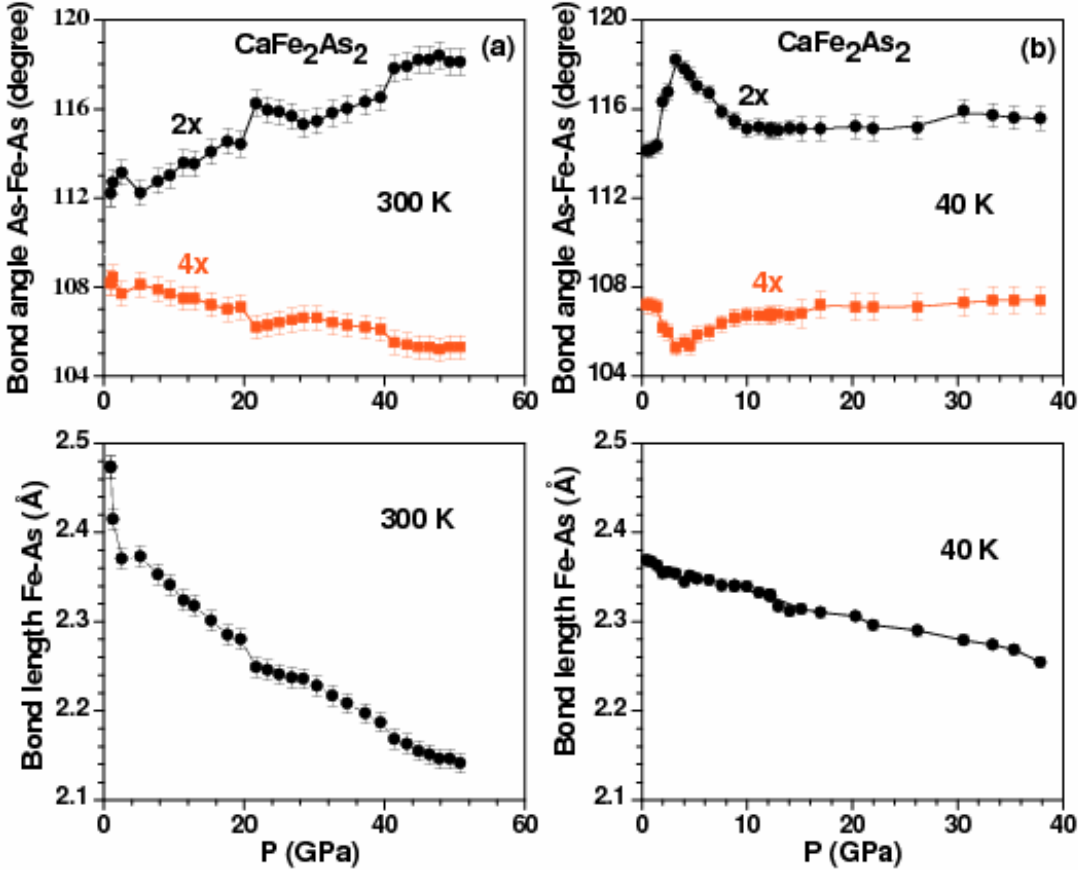